\documentclass[sigconf]{acmart}

\usepackage{flushend}
\usepackage{balance}

\usepackage{url}
\usepackage{rotating}
\usepackage{lipsum}
\usepackage{paralist}
\usepackage{multicol}
\usepackage{longtable}

\newif\ifremarks
\remarkstrue

\title{An Online Consent Maturity Model}

\subtitle{Moving from Acceptable Use towards Ethical Practice}

 \author{Vivien M. Rooney}
 \orcid{}
 \affiliation{
   \institution{IMT Atlantique, Lab-STICC,
  Universit\'e Bretagne Loire}
   \city{Rennes}   \country{France}}
 \email{vivrooney@gmail.com}
   
 \author{Simon N. Foley} 
 \orcid{0000-0002-0183-1215}
 \affiliation{
   \institution{IMT Atlantique, Lab-STICC,
  Universit\'e Bretagne Loire}
   \city{Rennes}   \country{France}}
 \email{simon.foley@imt-atlantique.fr}

\copyrightyear{2018} 
\acmYear{2018} 
\setcopyright{acmcopyright}
\acmConference[NSPW '18]{New Security Paradigms Workshop}{August 28--31, 2018}{Windsor, United Kingdom}
\acmBooktitle{New Security Paradigms Workshop (NSPW '18), August 28--31, 2018, Windsor, United Kingdom}
\acmDOI{10.1145/3285002.3285003}

\begin{document}

\begin{abstract}
Achieving informed consent in online and digital contexts is challenging for several reasons.  One reason is that conveying the meaning and implications of agreements to individuals is hindered by legalistic formats obscuring the potential harm that can ensue from analytics of data collected in a socio-technical context, such as online.  Furthermore, as technical capability advances, what can be achieved with data mining and initiatives outpaces statutory regulation, as well as the social norms that frame individual human understandings.  It is argued that the paradigm that currently underpins informed consent in online settings draws on ethical positions that are either utilitarian or legalistic.  In contrast, the adoption of an ethics of virtue approach as a new paradigm provides a framework for reconceptualising informed consent.  Characteristics that are material for informed consent, shared by Online Analytics and Qualitative Longitudinal Research, provide the inspiration and basis for this interdisciplinary approach, with the application of lessons learned in the practice and theory of one discipline to another.  


\end{abstract}


\maketitle

\section{Introduction}

Achieving informed consent in an online setting is the subject of scholarly interest \cite{christen2017, custers2016a}.  Among the difficulties identified with informed consent is its inability to meet the needs of the data giver.  This is evident, for instance, in the legalistic and complex manner that information is presented to the individual user.  The individual may skim the content, however, often they do not, and are motivated to press the `click through' button, often because of their desire to access a service.  They do not, however, want to have to negotiate their way through the detail of Acceptable Use Policies, or Privacy Policies, and even if they can do so, they are unlikely to understand the content \cite{bashir2015,Kelley:2009:NLP:1572532.1572538}.  The impenetrable quality of consent agreements is itself open to be exploited by data gatherers.  The legalistic language used reflects the power imbalance between parties to a consent agreement.  This manifests itself in the terms of an agreement, such as retention of data, its further use, and obscuring the interests of the data giver, such as the prerogative to withdraw their data.  Online data and meta-data is directly and indirectly gathered from users, analyzed and inferences are made; this activity is hereafter referred to as \textit{Online Analytics}.  
Online Analytics here refers to the analysis of data gathered in the context of research projects, as well as in contexts that are commercial. 

There is a distinction between a participant making a conscious choice to be involved in a research project, and the collection of personal data that can be analysed, particularly in relation to the awareness of the purpose for data collection.  However, it is the similarity between both, and what happens to the data collected in these two contexts that is salient for our purposes.  The conduct of Qualitative Longitudinal Research is one context for data collection.  This is where the process of data collection and analysis occurs over time. The second context for data collection is commercial purposes in an online setting, or data that has been collected neither explicitly for research nor commercial use, such as the data collected from a person receiving medical treatment in a hospital, that has been stored digitally, or has been converted to digital media subsequently.  The premise of what we argue is that what happens in Qualitative Longitudinal Research, the specific research context as understood in academia, has similarities with what is happening with other data, gathered for different purposes, not explicitly identified to people as being research, in the academic sense.  Examples of such contexts would be where people agree to the terms of an online service provider, where the uses that can be made of the data are similar to what can happen in Qualitative Longitudinal Research, although this is not made explicit, and indeed can be obscured.

Even if all the available information in a consent agreement is read and understood by an individual, the possible uses that may be made of data are likely to be conceptually abstract, at a remove from their reality and social norms, and hence disregarded.  Indeed it is difficult to convey the possible harm that can ensue from the practice of data analytics to an individual \cite{menlo:2012,menlocomp:2013}.  Information about potential harm is likely, therefore, to be disregarded by individuals.  

The argument being put forward is that the application of an interdisciplinary lens to what happens to data in online contexts, has shown that there are commonalities with Qualitative Longitudinal Research.  The commonalities particularly material to the practice of informed consent are our focus. The position argued is that Online Analytics is not only a technical activity, rather it is a social-science activity \cite{nuffield}.  Hence, the ethical norms of the latter should apply to the former.  On this basis, theory and practice of informed consent in Qualitative Longitudinal Research is a useful resource for the development of a new paradigm of consent in online contexts.

Drawing on interdisciplinary resources as a means of improving computer security is evident in recent research.  For instance, \cite{Pieters:nspw:2014} put forward the argument that the security of new technology could be improved by conceiving of its deployment as a social experiment.  In this way, potential hazards may be uncovered by being able to learn about the unexpected through observation.  This unpredictable quality of technology is the basis for Pacey's argument \cite{pacey:1990,pacey:1999} that values of justice and democracy should inform the design of new technology, as it is not possible to envisage the uses to which new technology will be put, as the social context for their deployment evolves in tandem with technological capabilities.  This is relevant to informed consent, as the unanticipated uses that can now be made of automated data emerge.  Online analytics is a crucial area where the social aspects of technology and informed consent intersect.
Ethically, how informed consent is currently conceptualised in online contexts is underpinned by either a utilitarian or legalistic approach, neither of which consider the needs of the individual user.  Firstly, consider the utilitarian approach.  For existing data sets, consent may be historic, and conducting data initiatives (DI) through further analysis of existing data, can be seen as a valuable resource, particularly in health care settings.  The absence of specific informed consent from the individuals concerned is a barrier to this secondary analysis, and obtaining consent for a new use of the data is considered to be impractical \cite{menlo:2012,menlocomp:2013} and an onerous imposition, because of the need to contact the possibly large number of individuals who gave their data in the past, consenting at that time, and impossible following the decease of participants.  Nevertheless, the ethical acceptability of a DI can be argued for on the basis that the data sets are a resource to be exploited, and that the ensuing benefits can accrue to the general population.  With this approach, the rationale for using data in ways absent specific consent for a new use of the data, as a DI would be, is that the ends justifies the means, or the greatest good acccrues to the greatest number of people.  This example encapsulates a utilitarian ethical approach to informed consent.

The second ethical approach, a legalistic one, in an online consent situation would be where the consent agreement is so complex, legally obscure and lengthy, that the individual cannot reasonably be expected to grasp its meaning.  While an individual may click `I Agree', their consent is not meaningfully informed.  However, while not wholly grasping what has been agreed, the act of consent suffices to shield a data gatherer from legal sanction if and when the data is used in ways that were obscured from, and may be an unwelcome surprise to, the individual as a function of the legalistic presentation of the agreement. While the agreement may meet statutory obligations, and therefore protect a data gatherer from legal sanction, the consequences for an individual can be unexpected and lead to unfair or harmful consequences.  

What is common to these ethical approaches underpinning informed consent is that they are both reactive.  As such, the utilitarian approach might be applied to the use of data that has been gathered historically, and decisions about what is acceptable in terms of its use are made from a reactive position.  The legalistic ethical approach makes decisions in light of the legal frameworks applicable, as well as the technological capabilities that are material to the activities of the data gatherer.  A consent agreement is written and rewritten with a view to protecting those activities from legal sanction.  The reactive quality is that the response to statutory regulation proceeds from awareness about liability based on current or anticipated activity that would exploit advances in technology.

We argue that an Ethics of Virtue approach to consent in online contexts would differ fundamentally to the reactive utilitarian and legalistic approaches above, by being a proactive framework.  As such, this constitutes a new paradigm.  An Ethics of Virtue approach to informed consent would impose a way of thinking about data that foregrounds the needs of the individual data giver.  We deem this approach to be a user-centred way of conceptualising informed consent in online contexts.

An Ethics of Virtue approach to consent means that rather than weighing up decisions on the basis of the greatest good for the greatest number, the utilitarian approach, or adhering to the letter of the law, the legalistic approach, that instead, when considering any decision about data use, that priority would be given to the needs of the individual data giver.  We argue that the reactive ethical frameworks, utilitarian and legalistic, can and ought to be superseded by a proactive Ethics of Virtue approach.  
To explicate our position, how and why informed consent has evolved to an Ethics of Virtue approach in a particular research area, Qualitative Longitudinal Research will be discussed.  In light of the material commonalities between Qualitative Longitudinal Research and Online Analytics, such as the social practices that both involve, insights gained in relation to informed consent through the practice and theory of Qualitative Longitudinal Research, will be the basis for reconceptualising informed consent in online contexts to adopt a prospective Ethics of Virtue approach. 
Furthermore, 
such an approach can be viewed as a human centred informed consent.  The utilitarian and legalistic ethical approaches, being reactive responses, are, we argue, reminiscent of design for use before use.  
Fischer~\cite{Fischer:2017,Fischer:2011} argued that 
designing systems in such a way that they inherently support the ability to be re-designed as requirements change or are better understood, that is, the paradigm of \textit{design for design after design}, coheres with a human-centred approach.  
Building on this and its interpretation in user-centred security \cite{Pieczul:2017:nspw}, we propose that if online consent were to be underpinned by an Ethics of Virtue approach, it would meet the need for a human centred informed consent, and provide for a fundamental change in how we conceptualise and develop practice in the area.

This article is organised as follows. 
Following the Introduction, Section~\ref{analogy} discusses the commonalities between Qualitative Longitudinal Research and Online Analytics, specifically what is material to informed consent.  
Section~\ref{ethics} introduces three ethical approaches which can underpin informed consent. 
Section~\ref{maturity} discusses the approaches to consent and related work, 
with examples illustrating how the maturity of consent can be judged. 
Section~\ref{s:paradigm} explores the new paradigm---an Ethics of Virtue approach to consent---while Section~\ref{s:tech} considers some of the requirements for a socio-technical system that might support it in practice. 

Informed consent here refers to the process of consent being requested and agreed to in online settings, however, we envisage that our position on informed consent applies to the creation and processing of digital data, or data transferred to digital format.

\section{Two views of social research} \label{analogy}


\subsection{Qualitative Longitudinal Research} 

In the practice of Qualitative Longitudinal Research the particular characteristics of how the research is conducted have resulted in a body of theory and practice on informed consent.  Psychologists and social scientists studying human behaviour and the social world have gathered information from, and about, ordinary people.  A core issue, based on the nature of the research, is that new ethical dilemmas arise in the field.  Researchers are confronted with situations that prompt the question `what do I do now?' \cite{ellis:2007}.  The characteristic novelty of such dilemmas means that ethical guidelines cannot provide a comprehensive set of specific answers that researchers can consult.  Rather than being able to rely on a codification of applicable rules providing answers, instead, the researcher themselves must confront, conceptualise and deal with each new dilemma.  The tools of self scrutiny and ongoing reflexivity \cite{guillemin:2004,tillmann:2003} throughout the research process, and in ethically important moments, are a means of achieving ethical practice.  With the ethical dilemmas that occur in the practice of Qualitative Longitudinal Research, there is ongoing refinement of theory and practice around informed consent.  An ethos of ethical practice underpins the mature approach to informed consent in Qualitative Longitudinal Research.  

\subsubsection{Example of Qualitative Longitudinal Research}

Qualitative Longitudinal Research explores human experience.  Methodologically diverse and adaptable, the challenges expected and encountered in fieldwork provide an opportunity for methodological innovation, including the practice of informed consent.  

In one project, Qualitative Longitudinal Research was used to explore the experience of people using communication technology in the context of platonic intimate relationships at a geographic distance \cite{rooney:2012}.   The technique of semi-structured interviewing was used to gather data, working with a cohort of 14 participants, with each person interviewed 3 or 4 times at intervals, over a period of 18 months.  Interviews were conducted using an interview schedule that was adapted during the course of the project, to reflect what was emerging in data analysis.  The interviews were audio recorded and transcribed in full.  The transcriptions were subject to analysis, such as the Grounded Theory techniques of line-by-line coding, focus coding and Memo writing \cite{charmaz:2006}.  The audio recordings were retained, and subsequent access was for clarification, as well as facilitating analysis.  
Throughout the iterative analytic process, data were subject to intense and repeated scrutiny.  During this process, inferences were drawn about the personal lives and social world of the participants, and emerging analysis was interpreted theoretically.

\subsubsection{Informed Consent and Qualitative Longitudinal Research}

For the current purpose of considering informed consent, particular characteristics of Qualitative Longitudinal Research are material.  This particular type of research is often exploratory, hence the direction or focus that analysis might take is unknown at the outset.  Consequently, the scope and focus of data collection can alter during the course of a project, as can data analysis.  A further complication is that data collection and analysis can take place in parallel, and influence each other, as for instance, is common with the use of a Grounded Theory approach \cite{charmaz:2006}.  In this way, therefore, the outcome of a research project is unknown at the outset.  With uncertainty of analytic direction, procedure and outcome, it follows that potential harms consequent to the research process are also uncertain.

Qualitative researchers are aware of the potential for harm, yet conveying this, as part of informed consent, to those considering becoming participants and to existing participants, is challenging.  It is typical that potential participants are enthusiastic to take part in the research, and also typical that they do not wish to hear detail about the process or the potential consequences \cite {wiles:2007}.  Indeed, the amount of information available may be overwhelming, and, even if it is listened to, or read by potential participants, the detail and broader implications of the content may not be grasped \cite{Rooney2015}.  The subject matter of the research can be deeply personal, and what is discussed can be personally meaningful for participants.  With semi-structured interviewing, what is discussed can be unpredictable.  Examples of what arose in the example above are seminal events, such as the death of family members, illnessness, and interpersonal conflicts.  The information disclosed to researchers is retained, and subject to analysis iteratively, hence analytic direction can change.  Both the substance of research, and the procedure for its conduct, have serious consequences.  When decisions are being made concerning the procedure for conducting research, consent is a key issue, providing for an exception to what otherwise may be deemed a violation of social norms, a civil tort, or even a criminal offence \cite{hurd:1996}. What takes place can impinge on participants and others in the short and long term.  Consequences may be at the personal or at the societal level.  Examples are damaged trust in interpersonal relationships, following publication of ethnographic research \cite{ellis:2007}, or the reproduction of racism \cite{rusert2009}.  



\subsection{Online Analytics} 

The practice of Social Research involves obtaining, retaining and analysing information about people.  Information that is personal, and often sensitive, is revealed by people about themselves. The desire to use a service may be underpinned by social or professional reasons, hence people are both willing and keen to engage with the service provider.  As such, they can be said to choose to give information, in the sense that they can be deemed legally to have an awareness of the scope of the clauses in an informed consent agreement, although often this is not read.  Further data collection can be opportunistic, steered by the earlier data collection and analysis.  As this occurs, such information can be linked to previous occasions of data gathering and analysis, and made sense of in the context of our cultural and social world.  Inferences are drawn about people based on scrutiny of what has been disclosed in conjunction with what has been retained.  Uncertainty of research outcome and consequences is intrinsic.  The preceeding describes what happens in both Qualitative Longitudinal Research and Online Analytics.  Online, a person's activity constitutes a trail, and this trail can be logged.  Retained, this record constitutes data that can be subject to analysis.  Over time, data continues to be compiled, accumulated and subject to further analysis. Data Initiatives \cite{nuffield}, where analysis of data sets can be conducted that was not envisaged, or even possible, at the time of initial data collection, is similar to the unknown outcome of qualitative research, where analysis that was not envisaged at the time of data collection, may be proposed subsequently.  
The technological ability to embrace complexity \cite{nuffield} means that data sets can be linked.  This provides opportunities for further analysis in ways that may not have been anticipated when a person's trail of activity began to be logged.  Over time, the accumulated data and analysis can be the basis for inferences being drawn about people.  What emerges may prompt further levels of analysis, and new inferences.  The outcome of this process of data logging, collection, retention, linking and analysis is unknown.  The consequences and implications of these procedures are obscure, particularly when data are held for long periods of time \cite{nuffield}.

Consequences could include the erosion of privacy resulting from the capabilities of technological advances, and ensuing harm.  An example is what can be inferred from linking online activity, physical location activity, in conjunction with, for example, medical records or purchasing records.  Taken together, the consequence could be that data gatherers would have greater knowledge about a person's vulnerability to a particular illness, than individuals themselves, or their family members, have. A consequence unlikely to be foreseen by individuals is such information being accessed and subject to algorithms to differentiate the cost of medical or life insurance offered, or even whether it is made available to them, or the availability of other financial products.  As the advances in technology illustrate, the challenge is ensuring consent that is informed. Conveying the risks involved is challenging \cite{menlo:2012}.  Potential harms are unclear.   
Even with data that is not deemed to be sensitive per se, data analysis can result in consequences that may be personal.  Information can be included as data, retained, and interpreted by a researcher.  Seemingly innocuous information can be retained and collated with existing data sets in ways that are unexpected by participants.  The availability of such information to the general public or to commercial interests, can have unforeseen consequences.  For instance, collecting information on an antivirus software preference could have implications that may not be apparent when such information was sought.  A response recorded that a preference was for `none' could have the potential for legal issues subsequently.  In light of such possibilities, the ideal is to foster awareness of the potential consequences of data analysis, and thereby enable participants to make informed decisions.  
What is clear is that the activity taking place has consequences that are social.  Further examples of where discrimination can occur, based on what is inferred from data analysis, are credit, housing provision and personnel selection \cite{romei2013}. As such, data analytics cannot be regarded as being solely a technical activity, rather it needs to be regarded differently, as a social practice.  Therefore, taking this view of Online Analytics, the ethical approach that underpins the concept of informed consent merits scrutiny from a social research perspective.  


\subsection{Shared characteristics}

We have outlined Qualitative Longitudinal Research and Online Analytics for the purpose of considering ethical practice and informed consent.  The following summarises their shared characteristics.  

\begin{itemize}
\item enthusiasm to sign up/take part
\item lack of enthusiasm for detail
\item individuals freely give information
\item personal and sensitive topic
\item longitudinal data collection
\item data retention
\item iterative data analysis and collection 
\item opportunistic data collection
\item unforeseen analysis can occur
\item inferences drawn about people
\item unforeseen harm is possible
\end{itemize}


The practice of Qualitative Longitudinal Research has specific methodological techniques and associated theory.  In general, cohorts of participants are numerically few, and there is individual, often personal, contact between researcher and participant.  This differs from research based on online activity that is used to generate statistical models.  In general, participants are numerous, and there is an absence of personal interaction.  However, the application of the body of knowledge that has developed around informed consent in the practice of Qualitative Longitudinal Research can, we believe, be informative for informed consent in the practice of Online Analytics. This is because some issues pertaining to informed consent, such as consent being a process rather than a once-off step, are thrown into sharp relief by the particular characteristics of qualitative longitudinal research.  The goal of using practice and theory underpinning Qualitative Longitudinal Research is to provide a means of enabling us to draw out some of these underlying issues, and illustrate their relevance and applicability to informed consent in an interdisciplinary context.

\section{Ethical Approaches to  Consent} \label{ethics}


Informed consent has evolved over time.  Rooted in the post World War II context, a biomedical model of informed consent emerged, and this was subsequently adopted and adapted for behavioural research \cite{menlo:2012,menlocomp:2013}. Practical issues in the field revealed the biomedical model of consent to be inadequate to meet the needs of both researchers and participants.  Conducting Qualitative Longitudinal Research, practitioners have continued to develop the theory and practice of informed consent.  As the models underpinned by utilitarian and legalistic ethical approaches proved inadequate, informed consent has evolved, and is underpinned in Qualitative Longitudinal Research by an Ethics of Virtue approach. This provides a framework that is coherent with a participant centred ethos.  Given that Qualitative Longitudinal Research, as well as Online Analytics, are both argued to constitute social practices, it is this ethical framework that is appropriate for a user centered approach to consent in online contexts.


Given the gravity and extent of the consequences for those involved, both the substance and procedure of research raise issues concerning what is deemed to be acceptable.  An ethical framework can provide a means of helping to judge what is right and what is wrong.  Three ethical approaches can provide a yardstick with which to judge consent.  These are utilitarian ethics, legalistic ethics and ethics of virtue.  

\paragraph{The utilitarian approach.}
The utilitarian approach to ethics is that the best moral action is the one that results in the most benefit.  This is illustrated, for instance, by the dilemma of six starving shipwreck survivors afloat in a lifeboat whose decision is to eat one of their number in order that the remaining five may survive \cite{dudley:1884}.  
Hence, the end justifies the means encapsulates this approach. 

\paragraph{The legalistic approach.}
Adherence to rules determines how an action is judged, independent of the outcome.  A swimming competition provides an example.  The outcome is appealed by the runner-up, such that the winner is disqualified.  This decision rests on a rule specifying that the winner would be the person whose \textit{hands} first touch the wall.  The disqualified winner had only one arm \cite{Hutchinson:1988}.  While rules are applied with precision, the outcome is unjust.

\paragraph{An ethics of virtue approach.}
This is a contextual approach to ethics, aspiring to ideals such as human dignity and autonomy.  The basis is personal integrity, with moral values internalised, and an emphasis on ethical intuitions, feelings and skills \cite{Macfarlane2009}.  Decision making relies less on weighing up an outcome, as the utilitarian approach would, or on rules to be applied, as a legalistic approach would.  Rather the aspiration is to proceed while foregrounding virtue.   An example is where a researcher makes contact with participants to revisit consent as data analysis proceeds \cite{Rooney2015} despite there being no formal requirement to do so.

\section{An Online Consent Maturity Model} \label{maturity}

As noted, consent in Qualitative Longitudinal Research has developed in light of the particular challenges that have emerged in practice.  The strategies and techniques that researchers have developed for ensuring that consent is truly informed are discussed below.  Online consent is also discussed in relation to the commonalities material to informed consent, shared with Qualitative Longitudinal Research.  The three ethical approaches outlined above are the framework by which the maturity of consent is judged and are summarised in Tables~\ref{CMC1} and~\ref{CMC2}.


\begin{table*}[htb]
\begin{center}
\begin{tabular}{|p{0.15\textwidth}| p{0.15\textwidth} || p{0.2\textwidth} | p{0.2\textwidth} | p{0.2\textwidth} |}\hline
&& \multicolumn{3}{|c|}{Consent Maturity Levels } \\ \cline{3-5}
Criteria & Description & Utilitarian & Legalistic & Virtue \\ 
\hline \hline
Initial Consent 		& Agreement to participate	
				& Opt-out 			
				& Opt-in 	
				& Opt-in, includes cooling-off period, following which a confirmatory opt-in is required, otherwise refusal is presumed
\\ \hline 
Altering consent    	& Modification of the agreement  		
					& Not facilitated		
					& Possible, onus on participant to discern procedure 	
					& Offered regularly and easily achieved		
\\ \hline
Withdrawing consent    	& Revoke the agreement to participate	 	
					& Not facilitated 
					& Possible, onus on participant to discern procedure
					& Offered regularly and easily achieved
\\ \hline
Scope of initial consent    	& What is encompassed by any initial agreement  	 	
			& As broad as possible: interpreted inclusively		
			& Interpreted in terms of what is reasonable or in the interest of the common good		
			& Limited to what was conveyed initially: interpreted strictly, erring on the side of exclusion		
\\ \hline
Ongoing consent    	& Consent is a process throughout the duration of a project	 	
			& To be avoided	
			& Avoided, consent is revisited in a strictly formal manner, adhering to the provisions of any agreement	
			& Participants are routinely consulted, reminded of the options available to them, such as withdrawal.	 Changes in consent are easily achieved.	
\\ \hline
Further consent    	& Consent is sought for any further or secondary analysis, Data Initiatives. 	 	
			& Opt-out, covered by initial consent, and if necessitated by legal provisions, is presented in a way that requires agreement in order to access information/services.		
			& Opt-out, adheres to strict legal requirements.	
			& Opt-in; a lack of response from participants to a request for further consent is regarded as a refusal.		
\\ \hline
\end{tabular}
\end{center}
\caption{Consent maturity criteria: consent actions} \label{CMC1}
\end{table*}

\begin{table*}[htb] 
\begin{center} 
\begin{tabular}{|p{0.15\textwidth}| p{0.15\textwidth} || p{0.2\textwidth} | p{0.2\textwidth} | p{0.2\textwidth} |}\hline
&& \multicolumn{3}{|c|}{Consent Maturity Levels } \\ \cline{3-5}
Criteria & Description & Utilitarian & Legalistic & Virtue \\ 
\hline \hline
Explanation of project    	& Comprehensive and relevant information provided  	 	
			& Minimum provided at outset 		
			& Specified precisely in legalistic language		
			& Elucidated clearly and accessibly, including possible consequences, updated as necessary, questions invited.	
\\ \hline
Informing Participants    	& Participants are kept up to date about the project and informed of what emerges from data analysis 	 	
			& Not necessary.		
			& Not envisaged, to be excluded from any agreement with participants		
			& Communicated to participants in an accessible way, along with appraisal, including start and completion date of analysis.  Response is invited, and taken on board.	
\\ \hline
Withdrawing data    	& Partial or complete withdrawal of data  	 	
			& 	Not facilitated 
			& Possible, onus on participant to discern procedure
			& Offered regularly and easily achieved
\\ \hline
Reuse of existing data sets    	& Existing data is subject to secondary analysis, such as Data Initiatives	 	
			& Acceptable		
			& Allowable if deemed to be in the interest of the common good and/or in line with current social norms		
			& Not envisaged: once initial purpose is complete, further use requires consent `ab initio'		
\\ \hline
Linking of data sets    	& Data Sets can be linked for analytic purposes 	 	
			& Acceptable		
			& Allowable if deemed to be in the interest of the common good and/or in line with current social norms.		
			& Not envisaged: 	once initial purpose is complete, further use requires consent `ab initio'		
\\ \hline
Trust    	& Mutual trust is built and maintained through fostering openness and equity	 	
			& Not envisaged.		
			& Not envisaged, access to information available as required by legal agreement.		
			& An intrinsic part of the approach, including easy access to information regarding past data breaches, remedies implemented, a history of the organisation's activities.		

\\ \hline
Reflexivity    	& Engaging in ongoing critical self-analysis of the project to improve and adapt 	 	
			& Not considered	
			& To be avoided, as required by legal provisions		
			& Intrinsic to the approach, including challenges/difficulties with the substance and procedure of the project, decisions taken are subject to scrutiny by interested parties, stakeholders and participants  	
\\ \hline
\end{tabular}
\end{center}
\caption{Consent maturity criteria: data and relationships} \label{CMC2}
\end{table*}

\subsection{Utilitarian: Consent at Maturity Level 1}

\paragraph{Potential harm} Milgram's \cite{milgram1963} studies on obedience to authority conducted in the 1960s illustrate a utilitarian approach to consent.  Conducted as social psychological research, this well-known example illustrates how the potential harm to participants was not considered in the experimental design.  Participants were deceived, thinking they were taking part in a memory and learning test.  They were instructed to administer what they were told were electric shocks to another person.  This other person was located in a separate room, and while out of sight, they could be heard crying out when the supposed electric shocks were administered by the participant.  The increasing severity of the electric shocks being administered produced louder expressions of pain, followed by eventual silence.  In reality, the person crying out was an actor in league with those running the study, and no electric shocks were administered.  Society as a whole gains from the knowledge that the study produced.  However, the means by which this is achieved failed to take account of the potential consequences for participants.  Ethically, this approach would be unacceptable now in Qualitative Longitudinal Research.


\paragraph{Analysis of historical data sets absent consent} An example of a utilitarian approach to consent in an online setting would be Data Initiatives on medical databases.  This would be, for instance, the linking of phenotypic and genomic data, as discussed in \cite{nuffield} in the context of care.data and the 100k Genome Project.  Data has been compiled in the UK health system over many years, with General Practitioners and hopsitals, adopting electronic systems in dealing with clients \cite{nuffield}. National Registries compile information on specific diseases, such as cancer, and specialised agencies such as The Health and Social Care Information Centre (HSCIC) compile information on health and social care \cite{nuffield}.  More recently, the technological capability to analyse large data sets has resulted in a move to embrace complexity and link medical data with additional lifestyle, environmental and social data in order to understand disease and optimise treatment.  Patient advocacy groups support Data Initiatives, based on real and legitimate interests in the potential benefits.  One argument is that such data is a state asset and, as such, its value ought to be realised \cite{nuffield}, although the absence of informed consent from participants remains a barrier to data analysis.  Obtaining consent in retrospect \cite{nissenbaum2009} would be practically impossible.  However, if a Utilitarian perspective is taken, an argument based on the potential benefits to society as a whole could be a basis for overcoming this barrier.  A `high public interest' is regarded as an exception to the need for specific consent, and argued to be essential in public health epidemiological research.  With national registries, opt out is argued to be incompatible with their ability to produce correct conclusions \cite{casali2014}. Knowledge emerging from the Data Initiatives could result in benefits in terms of the treatment and prevention of illness.  Clearly, the outcome likely to result from such analyses are a strong argument in favour of adopting the means to achieve the end.


Consent that is broad is also advocated, meaning one-time consent, that is `once forever' such that future processing of data for research purposes is permissible.  This means a person deciding \cite{nuffield} 
`that they are willing to give undefined researchers unconditional and irrevocable permission to use the data they provide in perpetuity, in ways to be determined by others.'  The ability to withdraw consent is possible, however, obtaining re-consent is regarded as being practically unfeasible, time-consuming, administratively burdensome, expensive and intrusive on patients' lives \cite{casali2014}.


\paragraph{Interpreting an act of consent in light of evolving social norms} Other arguments for allowing Data Initiatives, or further data analysis, in the absence of consent emerge when an interpretative stance is taken on a restrictive initial consent that was given at the time of data collection.
There are two ways of interpreting that initial consent.  One interpretation is literal, meaning that the consent is restricted to what was specified at the time, which would likely remain a barrier to Data Initiatives, as such analysis of data could not have been foreseen at that time.  Another interpretation, however, is intentional \cite{custers2016}.  This means that consent is interpreted more broadly, based on what the participant may have intended at the time when consent was given, in whatever form was in use at that particular time and place, and in light of the applicable social norms.  As social norms change, the interpretation of what was intended by the participant's agreement to provide their data, knowing that the data was being stored in an automated system, could also change.  An example would be an individual registering with an National Health Service hospital in the UK as a way of being tracked in a hospital setting, but without medical information being attached, as such a practice was not available, or standard, when that registration occurred.  Clearly, the social norm applicable at that time differs to current practices, when registration in a hospital means that medical records are attached.  However, the act of the individual can be interpreted in light of changing social norms.  It could be argued that the person giving their data was allowing for its use in whatever way was technically possible at that time and place.  If so, and the capabilities of Big Data now allow for complexity to be embraced, then it would be possible to infer from the historic initial consent, if any, agreement to whatever data use and analysis that might be technically possible at that time, and into the future.

\paragraph{The practice of consent as constitutive of social norms} Were a utilitarian perspective to be taken on consent, or the interpretative stance outlined above taken, then Data Initiatives, or further data analysis absent consent, could become an acceptable practice.  This practice would then, in turn, become part of shaping our social norms in relation to what is acceptable for data analysis and consent.  Reasonable expectations in relation to personal or sensitive data in existing data sets could evolve such that the expectation of privacy would be diminished.  The situation is then one where what is asserted to be, becomes descriptive of what is \cite{nissenbaum2009}, and thus becomes the de facto situation.  An example of this is where Mark Zuckerberg, in talking about Facebook, asserted that \textit{``[..] people have really gotten comfortable not only sharing more information and different kinds, but more openly and with more people [..] That social norm is just something that has evolved over time"} (cited in \cite{shoreStein2015}).  An example of the effect of an assertion such as this, is hackers disseminating 2016 Olympic athletes' medical information.  Rather than provoking outrage, instead what was reported as headline news was one athlete, Simone Biles, stating they were not ashamed of their medical condition, and another, Venus Williams, stating that they had adhered to the required rules.  While the focus shifts away from the fact that this information was made public, toward the substance of the information, the social norm regarding what we can expect for personal automated data also shifts.  The possible consequences and harm that might ensue for individuals remains unknown.  

\subsection{Legalistic: Consent at Maturity Level 2}

\paragraph{Written information} Procedural steps providing evidence of consent at the outset of qualitative research are commonly in writing \cite{wiles2012}, although oral evidence is acceptable \cite{wiles:2007}.  There is, however, a difficulty in conveying the extensive and complex information that is needed for consent to be truly informed in both Qualitative Longitudinal Research and online. In both cases, people are often enthusiastic about taking part, yet unenthusiastic about the detail of what their participation might entail \cite{wiles2012, schermer2014}.  Daunting amounts of information can result in potential participants choosing not to continue, or for consent to occur without the documentation being read and understood.  The practical option, for qualitative researchers and online, is to make information available to people, so that the material can be read and absorbed at a pace that suits the individual.  In practice, it is often the case that signed consent forms have not been read and, even if they have, recollecting what has been signed is problematic \cite{bashir2015, Rooney2015}.

\paragraph{Legalistic format and content} Consent for the automation of personal information, for instance, in commercial or medical contexts, is often conceptualised as a coarse-grained procedural necessity.  Such a contractual and legal step can be implemented as a once-off act, or through ticking a box for an Acceptable Use Policy.  With this procedural approach, the document representing consent may be lengthy, complex, referencing clauses and subclauses.  The expansion of such agreements over time has been illustrated by Shore and Steinman \cite{shoreStein2015}.  What is usual is an additional mandatory step confirming that the document has been read and understood.  If fundamental contractual changes are brought to users attention subsequently, as a notification of material changes, this is typically another lengthy legal document.  There may be a specific period of time during which it is possible to opt out, prior to policy changes taking effect.  This means that taking no action represents acceptance of the terms, and thus is considered to constitute consent.


\paragraph{A legalistic model of consent} This model of consent is underpinned by a contractual and legal focus.  Referred to for resolution in the event of a dispute, a reliance on contractual terms reflects concerns associated with traditions other than qualitative research \cite{guillemin:2004}, such as the biomedical tradition \cite{Macfarlane2009, wiles2012,menlocomp:2013}.  Consequently, although the process of obtaining consent in qualitative research or online may adhere to the legally required rules, in practice, people may not fully grasp either the nature of the process or its scope \cite{bashir2015}.  The changes in contractual terms and Acceptable Use Policies are prompted by changes in technological capability and data regulation.  Data gatherers aim to cover possible scenarios to protect themselves from claims based on ensuing harm from contractual breaches.  

Proposing that consent to the collection and retention of personal data be meaningful and fair has been posited as a way forward \cite{bashir2015,Alan2011-WERTFT}.  Building on the Fair Transaction model of consent \cite{Alan2011-WERTFT}, \cite{schermer2014}, argue that a strengthening of informed consent from a legal perspective could result in undermining the effectiveness of the consent mechanism itself, and thereby have a detrimental effect on trust in data processing and privacy protection.  The difficulty they identify is the disconnect between legal theory, presupposing rationality, and reality, where people agree to most requests for consent without reading them \cite{schermer2014}.  Furthermore, the strengthening of consent legalistically does not take account of the data gatherer's needs.  What is mooted as a way forward is an implied consent based on activity, such as being on a website that uses cookies.  A caveat to this is the technical competence that the majority in a society has.  With the Fair Transaction model of data use, subjects must be able to rely on socially accepted standards for data processing, and consent must be explicitly requested for reuse of data.  Enforcement should focus on those behaving unfairly against data subjects by processing data for other purposes. This is reliant on social norms and what is reasonable, and is to be inferred. The above coheres with the position being taken in this paper that a legalistic consent agreement cannot meet the needs it seeks to address, as it cannot cover all eventualities.  We argue that this inability lies in the approach being both reactive in how it deals with data and data regulation, and, as such, being underpinned by a legalistic approach to ethics.  

\paragraph{Trust and consent} The sensitive and personal nature of the information being disclosed presents a complication for ethical practice both online and for Qualitative Longitudinal Research.  During data gathering, qualitative researchers apply their skills to elicit rich data, for instance, during in-depth interviews.  A relationship of trust is likely to develop between researcher and participant \cite{Rooney2015}. 
The challenge for maintaining ethical practice is that such bonds can be sidelined unilaterally by the researcher when expedient, despite the consequences for participants. Nevertheless, in terms of adherence to specified procedures, having evidence of informed consent safeguards qualitative researchers \cite{wiles:2007}.  Similarly, an individual can perceive that they are in a relationship of trust with an online commercial entity.  Over time, with repeated contact, the individual may  rely on the online entity in a way that connotes an interpersonal relationship. If the entity provides a platform/forum on which interpersonal relationships are supported, as for example, Facebook does, then trust is a significant issue, as it is intertwined with personally meaningful relationships.  Similar to qualitative researchers, online entities can unilaterally set aside perceived bonds of trust when expedient, as Facebook has been accused of doing in the wake of the Cambridge Analytica revelations.  What appears to an individual to be a breach of their trust, is likely to be legally legitimate under the contractual terms to which they have agreed.  The position that \cite{schermer2014} argued, that strengthening informed consent from a legal perspective could weaken the consent mechanism, and thereby have a detrimental effect on trust in data processing and privacy protection, is exemplified by the issues between Facebook and Cambridge Analytica.  The issue is that, in theory, information presented to a person would be read and understood, whereas the reality is that participants often consent without reading the information provided \cite{schermer2014}.  Some agreements, such as those in a clinical setting, may require participants to indicate their understanding, as the onus to show that a participant has understood what is being requested may be more onerous than in another context, such as online.  As also noted earlier, even if read, the implications of the consent agreement can be obscure to an individual.  

\paragraph{Social norms and the interpretation of reasonableness} The legal fora for resolving disputes is one where reasonableness informs the interpretation of contracts.  This applies also to reasonable expectations about the collection and use of data.  As an example, \cite{casali2014} calls for a balance between privacy and health rights to \textit{reasonably} address concerns.  Thus, how reasonableness is interpreted is an important question as its meaning is directly linked to social norms.  The process of social norms changing is relevant to this legalistic approach, just as it is relevant to utilitarian ethics, as discussed above.  Social norms play a part in how decisions are made across the variety of our interactions.  A norm may evolve and emerge in very specific circumstances.  How a social norm works is relative to \cite{nuffield}, for instance, institutional practices.  Social norms are also shaped, for example, by evolving technological capability, such as the capacity to collate and analyse data sets.  Big Data means that we can embrace complexity \cite{nuffield}, 
and as we begin to do so, this activity plays a part in creating the reasonable expectations that people can rely on concerning what they have agreed to, and how that is interpreted in the event of a dispute.  Thus, even if specific analytic practices are omitted from the process of consent, social norms play a part in what is considered to be reasonable.  The Menlo Report \cite{menlo:2012} uses an objective standard as the ideal, and the following extract from the report highlights the importance of the evolving nature of social norms: \textit{`Ideally, researcher actions are measured using the objective standard of a reasonable researcher, who exercises the knowledge, skills, attention, and judgment that the community requires of its members to protect their interests and the interests of others. As researchers gain a greater understanding of how to reason about and apply ethical principles, community norms and expectations about what is reasonable will evolve.'}  In the event of a dispute, and a subsequent adjudication, a practice hitherto considered unacceptable, may be deemed acceptable in light of contemporaneous evolving social norms in relation to what is reasonable.

\paragraph{Withdrawing consent} Being able to withdraw from the process at any time is part of the procedure of consent in Qualitative Longitudinal Research.  In practice, however, this can be misunderstood.  Despite extensive information being given, there is scope for misunderstanding, and a tendency for participants to be coerced and feel obliged to continue \cite{holland:2006}.  A participant withdrawing from a cohort is a daunting prospect for a qualitative researcher, with the ensuing loss of time and effort invested.  Hence, it is clear why there is a reluctance to draw attention to, or a tendency to obscure, this aspect of consent, once the research is underway \cite{Rooney2015}.  In both Qualitative Longitudinal Research and online, participants are at a disadvantage in terms of what aspect of consent is reiterated and what is not.  The desire to retain participants, and being in a position to manipulate how and what information is emphasised, presents another challenge for ethical practice.  This is an illustration of the imbalance of power that is a common characteristic of the researcher/participant relationship \cite{etherington:2007}.  In this instance, this is manifest in how knowledge and information about the terms that have been agreed can be used to disadvantage participants \cite{bashir2015} who may be inclined to withdraw from a project, yet may be unaware or uncertain that they are free to do so.  This is illustrated online by the lengthy and legalistic nature of Acceptable Use Policies and agreements to consent.  As discussed earlier, people tend not to want to read or hear detailed information when signing up online, and even if they do, the legalistic terms of the agreement can obscure information on how to withdraw.  Obscuring the ability and entitlement to withdraw consent, either partially or fully, ensures that participants are put at a disadvantage.  


The issue of ownership of data is linked directly to who can decide on its use.  Ownership of data can be regarded as a lens through which the subtleties around informed consent can be explored.  In qualitative longitudinal research, while ownership per se is not explicit in informed consent agreements, ownership is implicit in the control of data.  For example, the withdrawal of data from a research project is facilitated by the informed consent agreed between Researcher and Participant.  This can, however, be qualified, such as by the imposition of a specific time frame for the prerogative to withdraw being exercised.  This specific time frame facilitates researchers being in a position to conduct, finalise and report on data analysis.  Another issue illuminated by the issue of data ownership is the extent to which participants are entitled, or can choose, to be kept informed about the research project to which their data is attached.  In much research, participants are unaware that they can withdraw at any time, and the signing of a consent form is regarded as an irrevocable step.  Highlighting the prerogative to withdraw is part of the Ethics of Virtue approach to informed consent, while less so within other frameworks.

\paragraph{Modifying consent in a legal model} Procedures and rules, such as legislation, can be amended to remedy unfair outcomes that occur following their strict application, such as that illustrated above by the example of the swimming competition, and to reflect society's evolving values and norms.  However, such remedies are likely to take place, and have effect, after time has elapsed, during which damage has occurred.  Facebook and Google \cite{solon2016,shoreStein2015} amend privacy settings and rules, however, this is self protection in the face of technological and legal developments, rather than acting in the interests of the individual users of their service.  \cite{christen2017}, for instance, reports that individuals are only partially aware of the effects, and beginning to appreciate the erosion of social meanings and the frailty of traditional social norms in the digital domain.  One of the fundamental difficulties with a legalistic approach to ethics in online consent situations is that technological capability outpaces, not alone social values and norms, but also how these are expressed formally in legal provisions.  The legislative process is not nimble \cite{nuffield}.  Even if legal provisions are obeyed, and the gathering of data takes place in a manner that is strictly legal, fairness or trust can be the casualty.

\subsection{Ethics of Virtue: Consent at Maturity Level~3}

An ethics of virtue approach to consent means behaving with respect for the humanity of participants, where consent is a participatory process of ongoing negotiation and ethical engagement.  

\paragraph{The need for an ethical approach to consent} Qualitative researchers face challenges because of the nature of their work.  For instance, some live in the community they are researching \cite{ellis:2007,hicks:2002} and for others, the topic of research means that close bonds develop with participants \cite{Rooney2015}.  Such challenges, and similar ones involving research on sensitive topics, are well known \cite{Rooney2015,dickson:2007}, and are compounded with a longitudinal design \cite{ellis:2007,hicks:2002,warin:2011,Rooney2015,saldana:2003}.  This is because openness from participants is encouraged by researchers needing rich data to analyse \cite{Rooney2015} and what participants disclose under these circumstances is subject to analysis as data.  A dilemma is objectifying the lives of peoples that the researcher has come to care about over the course of their research \cite{hicks:2002}.  When faced with such dilemmas, researchers argue that procedures have little impact on the actual ethical conduct of research \cite{guillemin:2004,wiles2012} with the responsibility to do so falling on the researcher.  Procedural consent forms signed by participants have little import on what emerges from the qualitative analytic process \cite{ellis:1996,kvale:2009}. As discussed earlier, the direction that data analysis will take is unknown at the outset of the research process, as is the process of analytic interpretation \cite{etherington:2007}.  With uncertainty as an intrinsic characteristic of this type of research, the consequences of the process, such as potential harms, are not foreseeable.  The challenge for researchers is conveying this uncertainty to participants when their consent is being sought.  This is especially so, as it is common for people to be enthusiastic when invited to take part in social psychological research.  The corresponding lack of enthusiasm for the extensive information that needs to be conveyed, noted earlier, is compounded because the information itself is laden with disciplinary jargon.  Despite adherence to the procedural aspects of consent, the difficulty is conveying the potential for harm to an enthusiastic research participant consenting to a qualitative research project.  
This dilemma is not easily remedied.  However, by adopting an ethics of virtue approach to consent, the aim is that informed consent is a mutual agreement to engage in a participatory research process.  Such a relationship is not based on adhering to procedural steps, rather, the relationship is an ethical one \cite{ellis:2007} where dignity takes precedence.  

\paragraph{Reflexivity and Relational Ethics} Resources from the tradition of qualitative research, such as reflexivity \cite{guillemin:2004} and relational ethics \cite{ellis:2007} are tools that can enhance ethical practice.  Reflexivity means that the researcher scrutinises their own actions.  For instance, in ethically important moments when a dilemma arises, the question asked by a researcher of themselves is \textit{`What should I do now?'} \cite{ellis:2007}.  In the field, procedural ethics may not provide an answer, and the response to a dilemma lies with the researcher.  Examples of such dilemmas in Qualitative Longitudinal Research are a participant requesting that unspecified audio recorded material be excluded as data, or that their real name be used \cite{Rooney2015}.  As a resource for researchers when faced with such questions, reflexivity is a useful and practical technique \cite{tillmann:2003}.  Macfarlane's perspective is to adopt a positive encouragement to behave ethically, rather than focussing on prohibitions.  Thus, the researcher aims for virtues of courage, respectfulness, resoluteness, sincerity, and humility \cite{Macfarlane2009}.  In the context of ethnographic and autoethnographic research on intimate others, Ellis \cite{ellis:2007} focusses on relational ethics. This is described as the recognition and valuing of mutual respect, dignity and connectedness between researcher and researched.  Particularly relevant to longitudinal research, Ellis asks us to consider how we deal with changing relationships with those we research over time.  Being able to respond to a moral dilemma from an ethics of virtue approach provides a framework for ethical practice.  

\paragraph{Taking time to consider opting in as part of consent} An ethics of virtue approach needs to be manifest throughout all aspects of the research process.  At the outset of a project, as noted above, the amount of information provided to potential participants can be overwhelming.  One of the strategies adopted to facilitate ethical practice in Qualitative Longitudinal Research is to allow a period of time during which people are asked to consider their participation.  If they then wish to participate, they initiate contact with the researcher.  This would be following a meeting outlining the research design, the unknown outcome, and potential harm.  In an online setting (online-based  data/activity), or where data is being collected digitally (it can be subject to subsequent automated analysis) this step would mean that consent would be opt-in.  A mandatory period of time for reflection, following the initial request for consent, would be followed by the assumption of non consent in the absence of contact initiated by the potential participant.  The aim of this would be to alleviate any pressure on people to become and continue as research participants \cite{Iphofen2013}, which is often experienced \cite{Rooney2015,kaye2014,wiles:2007}.


\paragraph{Ongoing communication between researcher and participant} In light of the uncertainty of research direction \cite{holland:2006} suggest that ongoing communication with participants should take place.  This strategy is part of an ethics of virtue approach, facilitating the participatory research process.  As an example, consider the anonymity that is typical in Qualitative Longitudinal Research.  Participants, however, may wish to choose to claim their contribution by being identified \cite{grinyer2002,kvale:2009,parker:2005}.  An example of this arose in research on intimacy \cite{Rooney2015} where a participant repeatedly expressed a preference that their real name be used.  Discomfort \cite{ellis:2007} with using the person's real name prompted the ongoing communication suggested by \cite{holland:2006}.  Following a dialogue during which this issue was discussed in detail, the participant's choice was to opt for anonymity, in light of the potential for harmful consequences for their family. This illustrates the practical effectiveness of an ethics of virtue approach to consent, where empathy between researcher and participant informed the response to a concern.  Note that if a legalistic approach to ethics were taken, the consent given to use a real name would have sufficed to comply with procedural ethical requirements.  This would have sidelined the potential harm to people not directly involved as participants.  Communicating with participants when making decisions that involve their data fosters mutual trust with the researcher, and sheds light on the process of making decisions that are not clear cut, and where uncertainty exists and mistakes can be made. 
Communicating mistakes to participants has the benefit of supporting an ethical relationship. In the commercial arena, ethical behaviour as a policy is gaining traction, for instance, transparency is advocated by Accenture Labs \cite{accenture}. In online and digital settings, \cite{nuffield} suggests that past failings concerning data ought to be communicated to participants, on the basis that the consequence of doing so is that trust with data givers is fostered.  The admission of past failings, and how these have been remedied, demonstrates good faith.  Furthermore, when a researcher is open about past errors, the participant can make a truly informed decision on consent.  With this information, an individual could choose that consent could be limited, for instance, to a particular component of the request being made.   


\paragraph{Mutual trust} Dilemmas that arise during the research process can illustrate the mutual trust that has developed in a research project where bonds develop, and personal disclosure occurs.  An example is an audio recorded semi-structured interview, after which a participant asks the researcher to \textit{`leave that stuff out, would you?'} This referred to recorded dialogue that was material to the research topic, however, was subsequently considered by the participant as being too sensitive for inclusion as data.  The extent of this omission is left to the researcher, and carrying this out in accordance with the participant's wishes necessitates that a relationship of trust exist between them.  Despite empathy, trust and caring for participants \cite{Rooney2015}, there is a complication for a researcher, concerning where their loyalty lies \cite{ellis:2007}.  For example, the question of whether a researcher is loyal to the interpersonal trust that has developed, or to their role as a data collector, remains.  This reiterates the shortcomings of a legalistic approach to ethics underlying consent: clearly, in such a situation, the procedure of consent forms signed at the outset has little import for the specific nuances \cite{guillemin:2004} of the situation of an interview.  Empathy and caring about the person inform what happens.  An ethics of virtue approach ensures that being a researcher is secondary to the responsibility to acknowledge the connection with the participant as a person \cite{ellis:2007}. Thus, rather than adhering to a set of specific rules, which could confer a freedom to interpret the participant's request to exclude data in a minimal way, the ethics of virtue approach underpins the decision made.  Any disclosure that might cause harm, in the widest sense, is excluded.  The participant's needs are at the centre of how and what decisions are made. The characteristics of Qualitative Longitudinal Research mean that practitioners aim for consent to be a continuous process \cite{saldana:2003}.   As discussed above, with a power imbalance being common in the research relationship, ensuring that participants are aware of their prerogative to withdraw consent is often sidelined \cite{wiles:2007}, as is the ability to withdraw some or all of their data.  While participants are not powerless \cite{etherington:2007} it is, however, their personal information that is being gathered, recorded, stored and analysed.  In effect, they trust the researcher with sensitive data, and are vulnerable in the relationship.  An ethics of virtue approach provides a framework for a more equitable research relationship.  When a participant's right to withdraw is reiterated throughout the research process, some vulnerability shifts toward the researcher \cite{Rooney2015}.  The once-off procedural act of consent is sidelined in favour of a more balanced process.  This does not only make for a more equitable research relationship in terms of trust, it more closely resembles the ideal of consent from an ethics of virtue approach.

\paragraph{Researcher vulnerability: trust the participant} Adjusting the formal procedure of consent, by postponing the request for a signature on a consent form until the end of data gathering, is another strategy adopted by qualitative researchers \cite{Rooney2015}.  When the formality of written consent is approached thus, it is an opportunity to engage meaningfully with the participant, and there are several consequences. Obtaining written evidence of consent at the outset of the process, as is usual, places the burden of trust on a participant. A benefit of adjusting formal procedure is that the burden of trust shifts from a participant, toward a researcher.  This is trust in the participant not to withdraw the oral consent they have already given, even at this late stage in the process of longitudinal research.  This step also provides another meaningful opportunity of withdrawing from participation \cite{wiles:2007} and to exclude all, or some, data.  Crucially, at this stage of the process of consent, the request is that a participant affix their signature to a form.  This occasion provides a further opportunity for each decision taken during the research process to become open to the joint scrutiny of researcher and participant.  The researcher can discuss, explain and justify their decision or action.  From an Ethics of Virtue perspective, they can satisfy the expectations that they have of themselves \cite{Rooney2015} to foreground virtue throughout the process.  A further advantage of delaying a formal procedure until the end of data gathering is that making ethical decisions in the absence of written evidence of consent may facilitate considering the import and consequences of such decisions for participants.  In addition, with the lapse of time, participants will have become familiar with the research relationship, and have had time to absorb and comprehend the information made available at the outset of the process, which would help to alleviate the difficulty of information overload.  Participants have the opportunity to question the process of the research, its outcome, and so forth \cite{Rooney2015}.   This exemplifies an Ethics of Virtue approach, consent at a maturity level of 3, with the aim that both parties can be satisfied that decisions taken were optimal in any particular situation.

\section{A new paradigm of consent} \label{s:paradigm}

We have argued that drawing on the practice and theory of informed consent in Qualitative Longitudinal Research provides a framework for a new paradigm of informed consent online.  The following are value-based behaviours to ensure comformity with the paradigm: 
cede power;
act with empathy;
trust participants;
anticipate moral dilemmas, and 
act in good faith.

There is some overlap among the items listed, and even taking all of them into account, there is no claim that an objective and correct way of making decisions, or behaving, is attainable.  Rather, the outcome is one where decisions taken are contextual and the aim is for stakeholders to act such that the criteria of virtue are met.  This is a standard that stakeholders impose on themselves, and the ways of behaving outlined in this section are intended as a supportive and practical framework.

\paragraph{Act in good faith} There is an imposition on the data gatherer to behave ethically.  Furthermore, they need to be able to demonstrate clearly that they did do so.  This would apply when any decision is being taken with regard to the use of data, and regardless of whether or not an opportunistic exploitation is legally defensible.  Rather, the onus, and thus the social norm, would be that an ethics of virtue informs the decision process at each step.  In the case of a claimed breach, the burden of showing good faith, and thus an ethics of virtue approach, lies with the data gatherer.  Thus, rather than a presumption of innocence in the event of an accusation, there is a presumption of guilt on the accused party.  While this inversion of what is usual in legal adjudication may seem to be an unfair way to proceed, it is, nevertheless, a useful way to adjust the existing power imbalance between the parties, where the data gatherer has control and resources, and the data giver is, therefore, at a disadvantage. In terms of a sanction for breach of ethical practice, a punitive approach to damages would be appropriate, in view of the difficulty of identifying harm to individual data givers, in the short term, as well as the long term, that might result from any breach.  This step would also indicate a move away from a legalistic approach to informed consent, where adherence to statutory frameworks can shield wrongdoers from legal sanction, despite for instance, opportunistic exploitation of legal loopholes.

\paragraph{Trust participants} Initial consent would be opt-in, exercised following a period of time for reflection.  The participant prerogative to withdraw some, or all, data and/or consent would be revisited regularly.  This prerogative needs to be easily implemented by participants in order for it to be meaningful.  Aiming for an Ethics of Virtue approach to consent, a data collector's policy could also include a history of positive interactions between themselves and stakeholders.  This would provide additional information on which existing and potential participants could base decisions about requests for data collection or use, as well as uses that are outside what was originally envisaged at initial agreement.  This would require, for instance, identifying and notifying participants of any issues, errors or problems, and demonstrate endeavours to remedy and prevent recurrences.  Input from participants would be part of this ongoing dialogue, providing the opportunity for ongoing reflexivity on the part of all involved to inform the use of data in the present and the future.  

The result is that the requirement to trust is more equitably distributed among the parties, in line with an Ethics of Virtue approach, and that building and maintaining trust is itself regarded as an important part of the process.  \cite {ohm2010} argues that fostering trust is so important that its rules ought to be formalised, developed and documented, with a view to sanctions for violating those rules.  In practical terms, this also means that mutual trust could be based on a history of interactions.   These decisions taken would be documented, and are then a record of the interaction including its outcome, and subject to joint scrutiny.  Another benefit of this approach, for online and digital contexts, is that the concern that research could be stymied by participant apathy would be lessened were a participatory ethos to be adopted.  As participants would be legitimate stakeholders, the social norm of trust in research, and its benefits, would be fostered.  This approach would be similar to the reflexivity that qualitative researchers use to foreground participant needs and thereby foster mutual trust.

\paragraph{Cede power}  This would mean fostering regular communication with participants to make them aware, and/or involve them in decisions being made, about the use of data.  Data could be conceptualised as a jointly created valuable resource.  Taking the participatory ethos of the new paradigm further, stakeholders would engage with each other in an interactive and responsive manner.  Participants can have input into research, from its design, to the analysis of data, to what is reported and how it is reported.  This technique would help to adjust the unequal power relationship between data giver and gatherer.  

\paragraph{Anticipate moral dilemmas} The usefulness of the Ethics of Virtue approach to informed consent is that it is prospective in terms of data gathering and use.  This means that it is anticipated that there will be novel situations, along with their associated moral dilemmas.  This is a given in view of technological innovation and its capabilities for data analytics.  The onus on a data gatherer is to be prepared for the occurrence of such a situation, and to ensure that how the dilemma is conceptualised and a course of action is chosen, is achieved in light of the requirement to act from a position of virtue.  This could draw on, for instance, Relational Ethics espoused by \cite{ellis:2007}, where the priority is valuing mutual respect, dignity and connectedness between researcher and researched.  In practical terms, it would mean having procedures and structure in place for communicating with stakeholders, relevant experts, and a way of facilitating decision making that anticipates the need for discussion of important issues.  This would be a proactive approach to moral dilemmas as part of informed consent.

\paragraph{Act with empathy} Acting positively with empathy would require that the data gatherer would aim to understand how the outcome of a decision they are taking could be meaningful for a participant.  Furthermore, the active part is that they engage with the participant on this basis, discussing how a particular issue is understood by both parties, and thus develop a way forward together.  This differs from being empathic and making decisions unilaterally; and is also a means of ensuring a participatory ethos.

As the above outlines, an Ethics of Virtue approach to online consent would be where data collection, retention and use becomes an ongoing process involving all stakeholders.  Proposals for consent have incorporated aspects of this approach, and the following examples point to how the new paradigm could work in practice.  

A concrete example of moving toward an Ethics of Virtue model of consent in practice is the Danish National Birth Cohort study \cite{Olsen2012}.  Based on their experience, they recommend that contact with participants is maintained, because they believe that it is important to provide participants with information and updates about the study, even despite the practical barrier of the expense that doing so would entail.  While they acknowledge the financial burden of maintaining personal contact with a large number of participants, the value of this outlay is in the retention of participants.  At a minimum, Olsen \cite{Olsen2012} suggests a public website where research results are available to the participants in the study.


A similar perspective is taken in the context of consent in international collaborative rare disease research \cite{gainotti2016}, where it is advocated that consent procedures include clarity in explaining the research, in governance, ethical oversight, re-contact policies, privacy measures, withdrawal policy, and a commitment to inform participants of changes to them.  In line with an Ethics of Virtue approach, they propose that participants be allowed time to think, reflect on information received and ask questions, prior to formal consent being requested, and that potential foreseeable uses of data and samples be explained.  The potential for harm is to be explained, for example, that access to data might be allowed where different regulatory frameworks apply, the misuse of data, misconduct, and hacking.  As the efficacy of de-identification is not conclusive \cite{ohm2010}, the possibility that data will be accessed, shared and linked to other sets of information must be explained.  In addition, the unforseeability of the purpose and extent of further usage must be clear.  Casali \cite{casali2014} proposes that consent means that patients will be informed that their data/tissues will be used for future research, they will be informed about the conditions under which these will be stored, making the protection safeguards a part of their consent. Under this scenario, the right to deny the consent and to withdraw it at any time is retained. 
In exploring ethical issues in the collection, linking and use of data in biomedical research and health care, the Nuffield report \cite{nuffield} proposes that Data Initiatives, defined as the re-use of data in novel contexts and linking them with data from other sources, be thought of as social practices, and recommend a publicly statable governance policy, specifying the grounds for granting or refusing access to data sets on the basis of reasonableness.  Ethical design is advocated, requiring the reconciliation of values and interests in a coherent set of morally reasonable expectations.  These would need to be publicly statable and subject to appropriate governance.  Reflecting the processual nature of consent, they also suggest the possibility of a `living' ethics and governance framework that would reflect participant expectations \cite{nuffield}.  Furthermore, what is advocated for research projects is the wider use of both explicit and flexible ethics and governance frameworks, as well as increased participation by research subjects in their design and governance.  

According to Ann Cavoukian, Information and Privacy Commissioner of Ontario, Canada,  rather than privacy being assured solely by compliance with regulatory frameworks, instead, privacy assurance must ideally become an organization's default mode of operation \cite{kingJess2014}.  The privacy nutrition labels developed by Kelley \cite{Kelley:2010:SPN:1753326.1753561} feature the design of a visual representation of a privacy policy, a label, such that people can see with ease what it contains, and be able to compare one policy with another.  This also facilitates people being able to choose to opt out if they wish to do so.  To go beyond the state of the art \cite{grace2017} proposes that a user-centric privacy preservation technology must provide: a fine-grained control of an online service's usage of their data; analysis of whether user preferences match the privacy stance of a service; and an understanding of the implications of usage of a service with respect to user privacy.  \cite{custers2016a} proposes that informed consent ought to have an expiry date as a way of reducing data reuse outside the original purpose for which consent was given, and that renewal must follow the same procedure.  With the new possibilities and implications facilitated by the reproduction and transmission of online data outpacing social norms \cite{christen2017}, the proposed way forward is to focus on enhancing the autonomy of the individual user, in terms of data control in line with their values, fairness in data use, and the apportioning of responsibility appropriately. Christen \cite {christen2017} argues also for supporting researchers technically and legally, as well as their longer-term relationship with individual participants, involving discussions with other participants and researchers.  This reflects the new paradigm, where researchers can revisit issues with participants, and participant involvement in the research process is encouraged and facilitated. As the above illustrate, suggestions range from being aspirational \cite{kingJess2014} to being rooted in practical experience \cite{Olsen2012}, or both \cite{nuffield}.  The proposals can be encompassed within the approach to informed consent underpinned by an ethics of virtue approach.







\subsection{The labours of Hercules}


The Ethics of Virtue approach to informed consent presents significant practical challenges for the handling of participant data, while aiming to meet this ideal level of maturity.  For instance, the provision of a history of interactions, positive or negative, and maintaining contact with participants are onerous requirements, in light of the practicality of long-term data gathering and analysis, perhaps with a time span of many years.  Also to be considered is the burden this type of requirement places on participants.  While some people may welcome a requirement to provide of a period of time for them to reflect on choosing to consent, others may not wish to avail of this.  Similarly, some people may welcome ongoing contact, for instance in a small scale qualitative study \cite{Rooney2015}, or regard contact as an incentive to continue to supply data in a large scale epidemiological project \cite{Olsen2012}, others may wish contact, or their involvement, to cease following initial data gathering. 

The maturity model aims to tease out the detail of how informed consent is approached.  In doing so, the challenges associated with achieving an Ethics of Virtue approach in practice, means that this is an ideal.  The practical reality at present is that research is conducted under existing constraints.  Many of these constraints hark back to the development of informed consent over time, such as the need for prior consent in a research project.  The consequences of decisions taken during its development in the past remain as part of the present reality that data gatherers and data givers negotiate in practice \cite{pathDep}.  The goal of altering the status quo, to take the example of prior consent in writing, could be leveraged by incorporating a provision where written evidence of informed consent, once obtained at the outset, is revisited later at key stages.  In qualitative research this might be when data analysis is being finalised.  In commercial contexts, a key stage for revisiting written consent might be if a company were to partner with another, and wished to use the data they already have in a novel way within the new context of the partnership.  We see the ideal, an equitable relationship between data giver and data gatherer, between participant and researcher, as worthwhile and considered goal.

\section{Socio-technical systems of consent} \label{s:tech}

The previous sections frame consent in Online Analytics as a social practice, as Qualitative Longitudinal Research is.  In practice, consent is implemented via a combination of social and technical processes and the consent maturity model provides the criteria with which to assess the efficacy of these processes.
In this section we consider the new paradigm of informed consent informed by an ethics of virtue, from a more technical systems perspective. In particular, we explore some of the challenges to developing socio-technical systems that support consent in practice. 

\subsection{Consent by design}

Utilitarian and legalistic consent presume that the data gatherer is best positioned to design the consent protocol 
by virtue of their understanding of how they intend to use the giver's data. 
However, as argued in the previous sections, the data gatherer is often ignorant of the privacy needs of the data giver, either intentionally or otherwise, as a matter of convenience.  
On the other hand the data giver, while ultimately the decider about their own privacy needs, can be ignorant of the implication of how the gatherer may use their data.  
This gives rise to a \textit{symmetry of ignorance} \cite{Rittel:1984} around the consent between the data gatherer and the data giver. 
If consent is regarded as an agreement on requirements then this symmetry of ignorance in consent can be seen in the same light as symmetry of ignorance in requirements \cite{Pieczul:2017:nspw} between the producer-developer of an API and the consumer-developer of the API. 
The data gatherer is an expert in data analysis but can be ignorant of aspects of data giver's needs, while the data giver is the expert in their own sense of private-self, but can be ignorant about the potential for, and consequences of, data analysis. 

This symmetry of ignorance around consent is not limited solely to the modalities of data gatherer and data giver; it can also involve social norms, legal regulations, technology, and so forth. 
The challenge is to design a system where expertise and ignorance about consent is distributed across all these modalities. 
In user-centered design terms, Rittel \cite{Rittel:1984} characterizes this kind of challenge as a wicked problem.
Fischer \cite{Fischer:2017,Fischer:2011} argues that user-centered design techniques focus on understanding the requirements of the user and ``designing the system for use before use''. 
There is similar emphasis in utilitarian and legalistic consent: requirements for consent are established by designing the consent protocol before use, and then using it afterwards. 
While this might be enhanced by the application of user-centered security principles \cite{MEZ:2005} to better understand the user requirements for consent, the paradigm of design for use before use is limited to utilitarian and legalistic consent. 

This separation between design, and then use, is contrary to an ethics of virtue approach to consent where user privacy needs evolve, new technologies are exploited and social norms and regulations change.
This fits with Fischer's view that it is not possible to design systems that anticipate all uses in advance and that one must ``design to support design after design'' \cite{Fischer:2011}. 
Therefore, in order to support consent at maturity level 3, one must design the consent protocol to support design after design. 

\subsection{Consent as a process}

\begin{figure}[htb]
\centering
\includegraphics[width=0.45\textwidth]{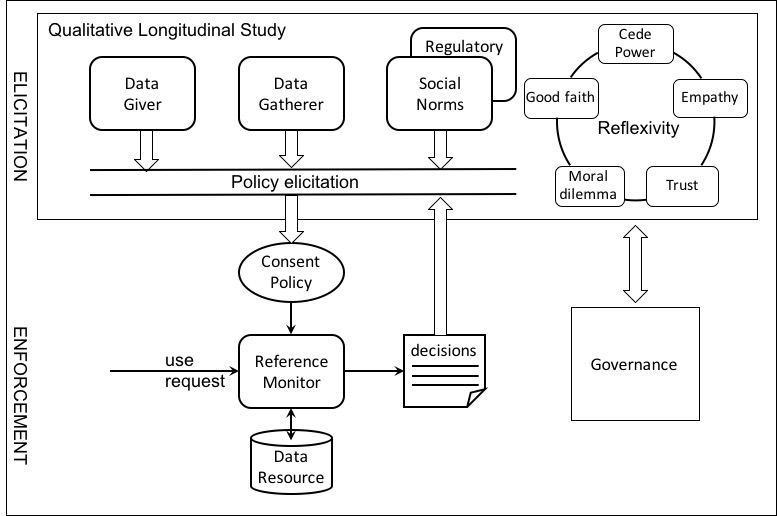}
\caption{Consent as Qualitative Longitudinal Research} \label{consentProcess}
\end{figure}

In the context of Online Analytics we identify two phases in assuring consent, namely the \textit{elicitation} of a consent policy related to 
the data of a giver and the subsequent \textit{enforcement} of that policy during any analysis carried out using the data.  
These Phases are depicted in Figure~\ref{consentProcess}.

The consent policy defines the conditions under which data of a giver may be used which we assume is enforced using a reference monitor.   
While ultimately constrained by the needs of the data giver, 
the policy should also reflect the needs of the data gatherer, 
the social norms and any regulation in place. 
At any moment in time, the consent policy gives a snapshot of the conditions under which user data may be accessed, and therefore, can be viewed as supporting a legalistic form of consent. 
However, the consent-policy is elicited via a qualitative \textit{longitudinal} study.
There is an ongoing process of reflexivity, 
re-visiting consent with the data giver, data gatherer and reflection on social norms and regulation, and past access decisions in order to refine the current consent policy. 
The value-based ways for a data gatherer to behave in order to fall within the new paradigm,
as discussed in the previous Section~\ref{s:paradigm}, are also part of this process, namely to act in good faith, to trust participants, anticipate moral dilemmas, act with empathy and to cede power.
It is this longitudinal reflexive nature of the consent process that enables 
the support of maturity level~3 consent.
We assume that all these activities are tracked and managed as part of 
broader data governance activity that can be used to provide the gatherer with assurance for the consent process.

\subsection{Consent elicitation}\label{s:elicit}

Our position is that consent elicitation is a Qualitative Longitudinal Research activity. 
A variety of Qualitative Research methods have been used previously to provide systematic means to elicit security and privacy requirements \cite{inglesant:2008,thomas:2014,secpre2017}.
For example, in \cite{inglesant:2008}, interviews and focus groups were carried out in order to understand Grid access control needs, and an access control language was developed to support these requirements. 
In \cite{awais:2016} Grounded Theory is used in conjunction with fault-trees as a methodological means to identify emergent threats.
Other studies have used an enthnomethodological approach to elicit privacy requirements for mobile applications \cite{thomas:2014}. 
Although these studies help elicit user needs, the difficulty lies in rendering them into policies that can be then enforced by a reference monitor.

Recognising that the codes, markups applied to encapsulate the meaning of text, uncovered during a Grounded Theory analysis of semi-structured interview data can be interpreted as policy attributes, a qualitative elicitation methodology \cite{secpre2017} has been developed whereby a Bayesian Network based policy can be systematically built from a (Grounded Theory) analysis of interview data. This policy represents a machine-interpretable encoding of the phenomena surrounding the user's needs.
We argue that an approach such as this can, in principle, be used during the elicitation phase for the consent policy.  

\begin{figure}[htb]
\includegraphics[width=0.475\textwidth]{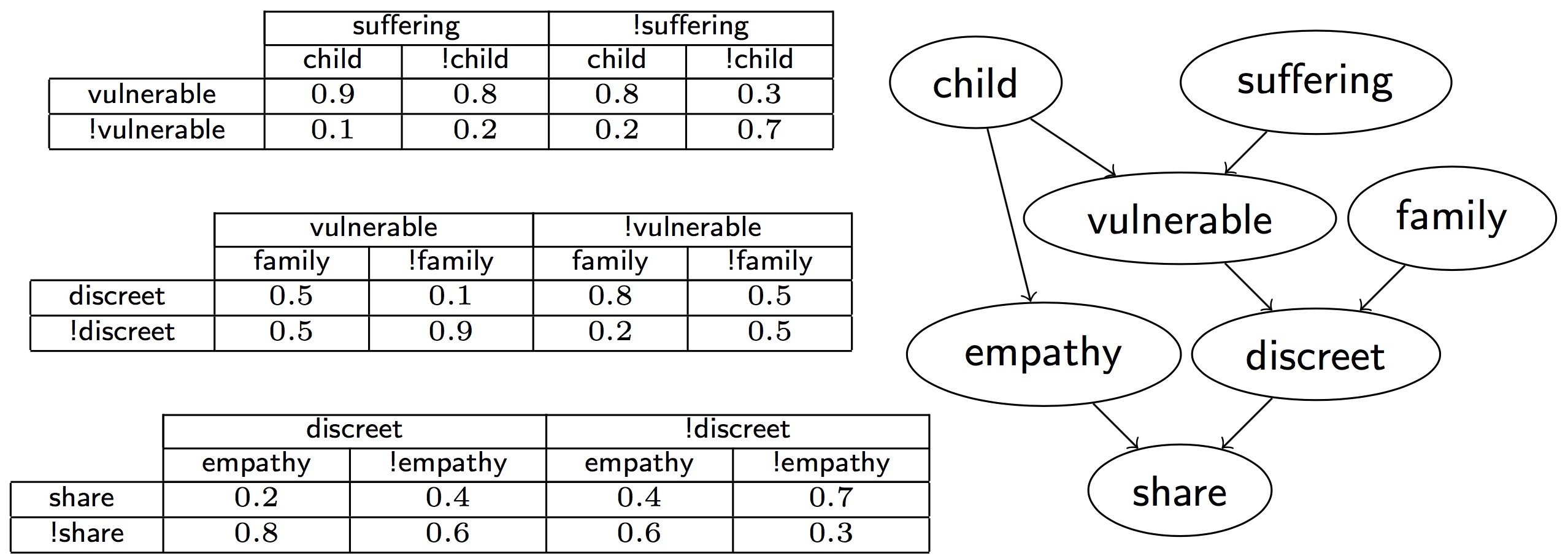}
\caption{Consent as user needs about sharing photographs} \label{BayesPol}
\end{figure}

For example, the policy in Figure~\ref{BayesPol} (from \cite{secpre2017}), elicited via a semi-structured interview, 
defines conditions under which the participant consents to sharing a photograph and 
identifies the phenomena that help to give meaning to the participant's sharing decisions. 
In this case, the participant feels that they are discreet if photograph sharing is limited to family members, 
and this is especially the case if the photograph has a vulnerable subject. 
Additionally, the participant feels that they would lack empathy if they were to share a photograph depicting a suffering subject.  
In sharing a photograph, the participant feels that this decision may generally indicate a lack of 
discretion or a lack of empathy, and this is reflected in the conditional probability table for \textsf{share} in Figure~\ref{BayesPol}.
That a Bayesian network is used as the policy is a reflection of the sometimes approximate nature of elicitation, and indeed consent.
In considering how their data might be shared or used, human users do not always communicate in terms of precise policy rules, rather, they are routinely discombobulated, vague and contradictory. 
Machine learning is used to help approximate the inevitable gaps in policy \cite{secpre2017}.
The original research \cite{secpre2017} was limited to eliciting the needs of the user as data giver.  However, the qualitative analysis can in principle be carried out across all modalities, including the needs of the data gatherer, while reflecting evolving social norms, regulations and so-forth.



\section{Conclusion}

Important issues in research, such as informed consent and intrusion, can be amplified in importance in qualitative longitudinal research.  \cite{holland:2006} point out that the extent to which consent is a process, rather than a single act, is exposed by the features of longitudinal research.  Having identified the commonalities between qualitative longitudinal research and what happens online, practices developed in the social sciences provide a useful resource when reconceptualising online consent. 

Moral values, such as beneficience, solidarity, justice, reciprocity, mutuality, citizenship and universality have emerged around the debate on consent in the last decade \cite{gainotti2016}. These values play a part in evolving social norms and, in turn, inform how data is conceptualised. What we think data is, and what can be done with data is, therefore, intrinsic to our social world.  An area of tension is that expectations based on social norms and moral values are relative, and thus subject to interpretation following a dispute.  Social norms are themselves being outstripped by the speed of technological innovation and capability; succinctly put, the law is not nimble enough to respond to social norms \cite{nuffield}, nor can it always reflect moral values \cite{Pieczul:2014:nspw}. In light of this, a rule based model of consent, a legalistic approach, is inadequate to meet the needs of stakeholders.  Taking an Ethics of Virtue approach, however, provides a way forward.  Advocating a positive encouragement to behave ethically, rather than a focus on what is prohibited \cite{Macfarlane2009}.  This would meet stakeholder needs, as concrete rules would not be definitive, given that they could not cover all possible scenarios, nor keep pace with technological developments.  Wiles \cite{wiles:2007} suggests that rather than forcing specific ethical decisions, that understanding moral codes and principles can, instead, be a resource in making sound and justifiable decisions. This approach could be put forward as the practice of ethics of virtue \cite{Macfarlane2009}, informed and guided by a principled approach.  The goal of fostering empathy, trust and developing an equitable ethos between stakeholders would be a resource to guide decision making when dilemmas arise, as they do in qualitative longitudinal research.  Practically and ideally, the specifics characteristic of principled and legalistic approaches to consent would be supplemented by a way of conceptualising data that foregrounds the human dignity and the relationships between stakeholders. The requirements for achieving an Ethics of Virtue approach to informed consent may appear onerous in view of, for instance, possible apathy from participants stymieing research.  The challenges of adhering to ethical practice may put researchers to the test, and the additional ensuing burden also stymie research.  However, another perspective is that the onus would be on the researcher to create an incentive, based on altruism, such that participants make the choice to opt-in.  There is evidence that if people understand how a behaviour contributes to shared goals for a relationship then they will be more accepting of a privacy threat \cite{horne2014}.  Similarly, understanding the beneficial goal of research when a request is being made could contribute to a willingness to be inconvenienced, and to consider participation.  



Despite the onerous nature of adopting informed consent based on an Ethics of Virtue approach, the need for such an approach is clear in the present situation, as fundamental changes around data analytics emerge with advances in technology.  Pacey argues \cite{pacey:1999,pacey:1990} that the design of new technology should be informed by ideals of justice and democracy.  This he contrasts with the technological imperative, where technological capability determines what we do, and how we use our abilities.  While an Ethics of Virtue approach to informed consent has aspirational qualities, the impact of our technological capability needs to be considered as being fundamental, rather than as an afterthought.

\subsection{Limitations and Future Work}

Our view is that informed consent is achieved through qualitative longitudinal research and we have
sketched one example of how this might be done in a systematic way in order to generate consent policies for data givers that could be enforced by a reference monitor/policy enforcement point.  
In practice, however, eliciting consent in this way, even for individual participants, is non-trivial due to the laborious and  time-consuming nature of Qualitative Longitudinal Research techniques such as Grounded Theory. 
Naively applied, it will not scale to large, or even small-sized user populations.  
In general Qualitative Longitudinal Research is conducted with numerically small cohorts of participants, whereas online and in digital contexts, the opposite applies.  Furthermore, in contrast to digital contexts of data analysis, qualitative data analysis can be a lengthy process. 
There are a number of ways that this challenge might be addressed. 

A contemporary approach could conduct the Qualitative Longitudinal Research with focus groups, representing different cross-sections of the data-giver population. 
An elicited policy is thus a reflection of the consent-related phenomena that are considered important by a 
cross-section of the community. 
As a Bayesian network, the policy can be individualised by tailoring the probabilistic variables to match user preferences, for example through user-questionnaires. 
Smart agents could further interact with the data-giver, to explore new ways of understanding consent with the user.
For example, in  \cite{secpre2017} we sketch a smartphone agent that learns/refines the photograph sharing policy based on  subsequent sharing decisions by the user that override their original policy.  To fully embrace the paradigm, elicitation would need to identify new phenomena in the users thinking over time.  

Looking to the future, the practical challenges for both Qualitative Longitudinal Research and informed consent underpinned by an Ethics of Virtue remain.  Locating the perspective on informed consent being put forward here in a broader contextual framework related to the morality of technology and information ethics, are possible directions for future work, as is the potential for incorporating the nuances of other ethical frameworks.
It is possible to envisage that in 2038, the data gatherers, social scientists and applied psychologists will spend their time studying consent and ensuring that it is well understood and well implemented.  
Just as there can be value in engaging Data Scientists in Online Analytics applications, so too should social scientists and applied psychologists be engaged to ensure that user needs in consent are addressed.   

\begin{acks}
We thank our anonymous reviewers, workshop attendees, and our shepherds Markus Christen and Lizzie Coles-Kemp, for their help in making this a better paper
This work was supported, in part, by Science Foundation Ireland grant SFI/12/RC/2289 and by the Cyber CNI Chair of Institute Mines-T\'el\'ecom which is held by IMT Atlantique and supported by Airbus Defence and Space, Amossys, BNP Parisbas, EDF, Orange, La Poste, Nokia, Soci\'et\'e G\'en\'erale and the Regional Council of Brittany; it has been acknowledged by the French Centre of Excellence in Cybersecurity.
\end{acks}

\bibliographystyle{ACM-Reference-Format}

\end{document}